# Towards Vulnerability Analysis of Voice-Driven Interfaces and Countermeasures for Replay Attacks

Khalid Mahmood Malik
School of Engineering and Computer Science
Oakland University
Rochester, MI, 48309, USA
mahmood@oakland.edu

Hafiz Malik
College of Engineering and Computer Science, University of Michigan-Dearborn
Dearborn, MI, 48128, USA
hafiz@umich.edu

Roland Baumann
School of Engineering and Computer Science
Oakland University
Rochester, MI, 48309, USA
rbaumann@oakland.edu

*Abstract*—Fake audio detection is expected to become an important research area in the field of smart speakers such as Google Home, Amazon Echo and chatbots developed for these platforms. This paper presents replay attack vulnerability of voice-driven interfaces and proposes a countermeasure to detect replay attack on these platforms. This paper presents a novel framework to model replay attack distortion, and then use a non-learning-based method for replay attack detection on smart speakers. The reply attack distortion is modeled as a higher-order nonlinearity in the replay attack audio. Higher-order spectral analysis (HOSA) is used to capture characteristics distortions in the replay audio. Effectiveness of the proposed countermeasure scheme is evaluated on original speech as well as corresponding replayed recordings. The replay attack recordings are successfully injected into the Google Home device via Amazon Alexa using the drop-in conferencing feature.

## I. INTRODUCTION

The growing trend of personalization, increasing number of smart homes, the desire for easy control of home IoT devices and rising consumer preference for luxurious entertainment systems are driving factors for the tremendous growth of smart speakers. Alone in 2017, approximately 40 million adults in the United States have adopted voice activated smart speakers. With a compound annual growth rate of 35%, global smart speaker market is projected to rise during forecast period 2017-2024 [1].
Gartner estimated that by 2020, 75% of US households are expected to have voice-driven interfaces, e.g., Alexa, Cortana, Google Assistant, Siri, and, so on [2], and worldwide spending on these platforms and devices is expected to be more than $3.5 billion by 2021 [3]. Currently, based on intelligent virtual assistant used in smart speakers, market leaders include Alexa, Cortana, Google Assistant, and Siri. Various home and office voice-driven applications rely on software development platforms provided by Alexa skills and Google Action. It is estimated that currently chatbots are handling around 30% of customer-service requests, and by 2020 chatbots are expected to handle 85% of customer-service interactions [4]. In addition, through voice, intelligent virtual assistants and smart speakers are being used to remotely control different Internet of things (IoT) gadgets such as controlling thermostats and doorlocks. Therefore, it becomes imperative to secure the voice-drive interfaces and associated applications and access control systems (speaker recognition systems) are vulnerable to replay audio (RA) attacks, impersonation, speech synthesis and voice conversion. Smart speakers enables attacker to remotely attack voice-driven interfaces and applications. Among these four attacks, audio replay is the easiest to exploit [6], whereby the pre-recorded speech of the target speaker is played back to for automatic speaker verification (ASV) task. The ASV, a key component of voice-based authentication and access control systems, is a process of the authenticating users by doing analysis on their speech utterances. The ASV has received significant attention in the last two decades due to its convenience, low cost, and remote operability with simple devices like mobile phones. The role of ASV is expected to increase further due to proliferation of voice-driven interfaces and virtual personal assistant-enabled wireless speakers.

Many technologies are used for ASV such as frequency eestimation, hidden Markov models, Gaussian mixture models, vector quantization, decision trees, and neural networks [5]. The ASVspoof 2017 Challenge [18] was focused on the exploiting shortcomings of existing state-of-the-art to detect replay attacks under diverse conditions. Efforts have been made to investigate replay attacks on ASV systems [12–17]. For instance, Patil et al. in [12] study the spectral changes due to the transmission and channel characteristic of replay devices for replay detection. Another attempt was made to capture the channel information embedded in the low signal to noise ratio region, a single frequency filtering feature with high spectro-temporal resolution was proposed in [13]. Most of the presented works in ASVspoof 2017 challenge used a combination of different features and classifiers to improve performance of replay detection system. The combined feature vector include constant Q cepstral coefficients, mel-frequency cepstral coefficients, linear frequency cepstral coefficients, rectangular filter cepstral coefficients, perceptual linear predictive and deep features as front-ends [14, 15]. Magnitude-based features are widely used in replay attack detection, [14, 15], and frequency modulation (FM) features have been used in speech recognition and speaker recognition [16, 17].

It has been demonstrated that replay attack introduces distortions in the spoofed speech [9,10]. Most existing state of the art methods mainly rely on machine learning based approaches. These approaches process the input speech signal for feature extraction that are used to train a classifier to learn the underlying distortion model. For example, [19] authors proposed light convolutional neural network (LCNN) classifier to extract high-level features from the log power spectrum, together with a Gaussian Mixture Model (GMM). GMM, support vector machine (SVM) and i-vector Gaussian probabilistic linear discriminant analysis were employed as back-end classifiers [20]. However, little work has been done on replay attack detection using higher-order spectral analysis (HOSA) features to capture traces of replay attack distortion and detection. Additionally, no work has been reported to study possibility of replay attacks and their countermeasures in smart speakers' environment.

This main contributions of this paper are:
1. This paper demonstrates that ASV feature of voice-driven interfaces, e.g., Google Home is vulnerable to replay attacks and thus all the skills and actions built on these platform, including many having critical financial data, could be exploited easily even by relatively less tech savvy impersonators.
2. According to best of our knowledge, there does not exist any attempt for the vulnerability analysis and exploitation of audio replay attack on Google home and Amazon Alexa.
3. We have modeled replay attack as a higher-order linearity beyond $6^{th}$-order (see Figure 2).
4. A countermeasure based on HOSA framework is proposed to detect reply attack.

## II. VULNERABILITY ANALYSIS OF REPLAY ATTACKS IN SMART SPEAKERS

This section describes vulnerability analysis and exploitation of audio replay to understand what is the performance of Automatic Speaker Verification (ASV) system used in the Amazon Echo and Google Home smart speakers We conducted several experiments to determine the capabilities of the ASV in these devices.

### A. Experiment 1 – Vulnerability Analysis of Replay Attacks in Amazon Echo

We have tested the ASV capabilities of the Amazon Echo by placing an order for small items such as candy, as non-owner of the device. Amazon Echo ASV was unable to assess who was placing the order. Similarly, we found that any person could use the Amazon Echo to turn on and off IoT connected lights in the home.

Despite the fact that all of the functions of the Amazon Echo are available to any user we did replay a recording of the device owner asking "Alexa, Who am I?". The Amazon Echo replied with the device owners name. This further proved that what limited voice recognition the device has, it is not capable of distinguishing recorded audio from a real voice.

### B. Experiment 2 – Vulnerability Analysis of Replay Attacks in Google Home

The Google Home device does use voice recognition to offer secure purchases and access control. However, our experiment revealed that the speaker verification is limited to authenticating the wake word, usually "OK Google". Once the wake word has been used to activate the device no further voice verification is performed on subsequent commands. This makes it possible for anyone that has a recording of the owner using the wake word to then have full access to the device.

To verify that only the initial wake word "OK Google" is checked, we took a recording of the account owner saying the wake word and then added in a completely different voice requesting to purchase something. At no point in time did the device question why a different voice was given to the device.

In another test, we took a recording of the male account owner saying, "OK Google" and then followed it up with a recording of a female voice saying, "Who am I?" the Google Home device responded with the male account owners name. It shows that Google Home performs voice verification only on the wake word.

### C. Experiment 3 – Introduction of multiple replays using Drop in conference feature of Amazon Echo.

Experiments 1 and 2 demonstrates that the speaker verification capabilities of the current generation of smart speaker devices are limited. While on the surface it appears that since smart speakers are located within the users home the damage to be done is limited to some mischief by people near the device. We considered if the capabilities of smart speakers could be exploited to unlock an IoT connected door system or change the settings on an IoT connected thermostat. To better understand the severity of audio replay attacks, consider a home that uses a Google Home device to control a door lock. All that would be required to unlock the doors of that home would be a replay of the owner using the wake word. This could be a genuine copy or a synthesized voice. Once the wake word is played and accepted, any voice could request the doors to be unlocked.

For this experiment we hid an Amazon Echo device behind a TV. We then used the Drop In Audio conferencing feature of the Amazon Alexa to replay voice recordings to the Google Home Speaker. We were able to replay a recording of "OK, Google, Turn on Office Lamp" via the audio conferenced Amazon Alexa from another home. The Google Home device did turn on and off the lights as requested in the replays.

One can envisage that while the Drop-In feature of the Amazon Echo made it very easy to perform this type of attack, it would be relatively easy to use other equipment to replay the attack into a person's home. For example, one can use a RaspberyPI equipped with an MP3 board to replay the required sounds to get the smart speaker to unlock the doors. For the Internet connectivity required to perpetuate this type of attack this could be done by knowing the home owners WiFi key or using a Cellular WiFi hotspot device.

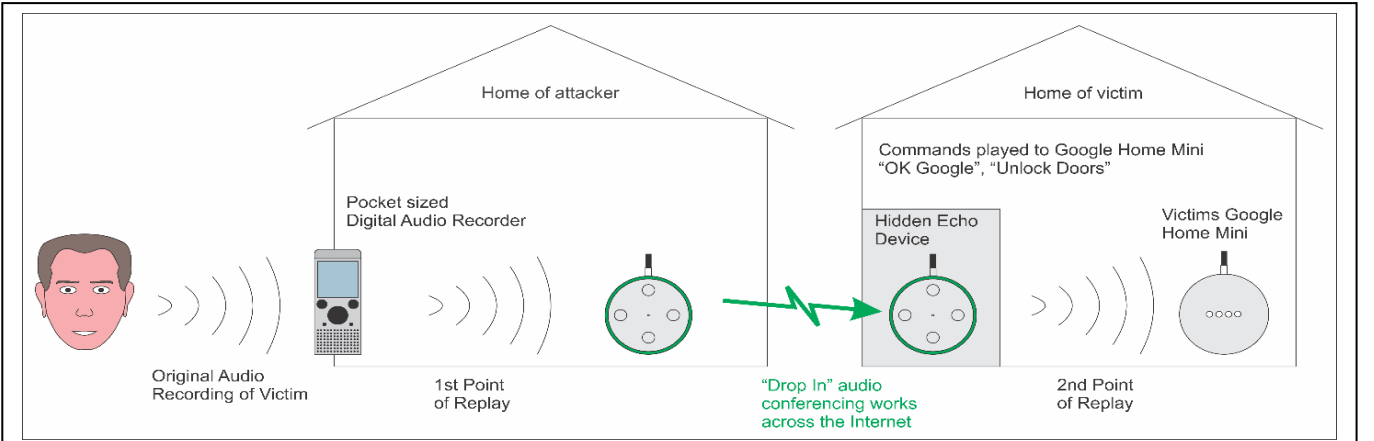

Figure 1: A scenario of Drop-in conferencing features of Echo device to generate replay attack on Google Home's ASV system

## III. REPLAY ATTACK MODELING

It has been demonstrated that replay attack introduces distortions in the spoofed speech [9,10]. Most of existing state of the art mainly rely on machine learning based methods. These approaches process the input speech signal for feature extraction that are used to train a classifier to learn the underlying distortion model. In this paper, we present a framework to model replay attack distortion, and then use a non-learning-based method for replay attack detection on smart speakers. As shown in Figure 2, the microphone and speaker are modeled as non-linear devices. The *Mic-Speaker-Mic* (*MSM*) processing chain of the replay attack, therefore, is expected to introduce nonlinearity in the resulting replay attack signal generated using proposed Alexa drop-in attack.

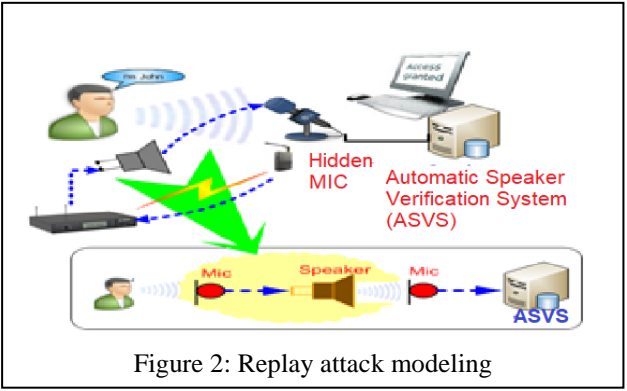

Figure 2: Replay attack modeling

## IV. REPLAY ATTACK DETECTION FRAMEWORK FOR SMART SPEAKERS

We propose to use higher-order spectral analysis (HOSA)-based features to capture traces of replay attack distortion and thus detect them. Details of the proposed approach is provided in the following subsections.

### A. HOSA-based detection:

The microphone/specific distortions such as harmonic–, intermodulation (IM)–, and difference-frequency (DF)– distortions. The presence of harmonic components at the output of a nonlinear system with pure tone input is called as harmonic distortion. System nonlinearity can cause IM distortion in the output when a complex signal (e.g., speech) is applied at the input of a nonlinear system. It causes the output signal to be sums and differences of the input signals fundamental frequencies and their harmonics, that is, $f_1 \pm f_2$, $f_2 \pm 2f_1$, $f_2 \pm 3f_1$, etc. Given a nonlinear system is excited with sum of sinusoids with same magnitudes then system nonlinearity can cause difference-frequency distortion at the output, e.g., $2f_2 - f_1$, $2f_1 - f_2$, $3f_1 - 2f_2$, etc.

It has been shown in [9] that microphone response can be approximated using following discrete time-invariant Hammerstein series model,

$$y[n] = \sum_k g_1[k]x[n-k] + \sum_\tau g_2[\tau]x[n-\tau]$$

The microphone (resp. speaker) nonlinearity introduces higher-order correlations at its output. The MSM processing chain, therefore, can be modeled using a higher-order nonlinear system. To capture it HOSA is used. Specifically, higher-order cumulants (resp. bicoherence) [8] is used to capture higher-order correlations. The bicoherence, $(f_1, f_1)$, of a signal $y[n]$ is a normalized version of 2-dimensional Fourier transform of the third-order cumulants, that is,

$$B(f_1, f_2) = \sum_{k_1,k_2=-\infty}^{\infty} \kappa_y^3(k_1, k_1) e^{-j2\pi(f_1 k_1 + f_2 k_2)}$$
$$= Y(2\pi f_1)Y(2\pi f_2)Y^*(2\pi f_1 + 2\pi f_1)$$

Here, $\kappa_y^3(k_1, k_1)$ denotes third-order cumulant of $y[n]$, and is defined as,

$$\kappa_y^3(k_1, k_1) = E\{y^*[n]y[n+k_1]y[n+k_2]\}$$

Here, $E\{.\}$ denotes expectation. Sometimes, it is more convenient to use the normalized value of the bispectrum which is also known as bicoherence. This bicoherence is given by the following equation [8],

$$B(f_1, f_2) = \frac{Y(2\pi f_1)Y(2\pi f_2)Y^*(2\pi f_1 + 2\pi f_1)}{|Y(2\pi f_1)Y(2\pi f_2)Y^*(2\pi f_1 + 2\pi f_1)|}$$

It is important to highlight impact of nonlinearity on bicoherence spectrum. Consider a pair of sinusoids with frequencies $f_1$ and $f_2$; the IM distortion will result in a new signal at $f_1 \pm f_2$ whose magnitude is correlated to $f_1$ and $f_2$, which will result in a high magnitude value in the bicoherence magnitude. Moreover, if the input sinusoids have phases, $\theta_1$ and $\theta_2$, then the phase of the nonlinearity induced intermodulation components $f_1 \pm f_2$ are $\theta_1 \pm \theta_2$. It is easy to see that the bicoherence has a zero phase and a bias towards $\pi/2$ may also occur due to harmonic auto-correlations. In general, the average bicoherence magnitude would increase as the amount of quadratic phase-coupling (QPC) grows. It can be concluded that a replay attack is expected to: (i) increase in the magnitude of bicoherence for certain harmonics, and (ii) the phase of bicoherence bias towards 0 and/or $\pi/2$ at IM distortion frequencies.

To capture traces of a replay attack, intermodulation distortion, QPC, Gaussianity test statistics, and linearity statistics can be used. For this paper, QPC, Gaussianity test statistics, and linearity statistics are used. The motivation behind focusing on intermodulation distortion is that it is more dominant in the cloned signal. To verify this claim, we estimated the bicoherence from both the speech and the corresponding cloned recordings. Shown in the left panel of Fig. 3 is the bicoherence magnitude plot of an audio recording and in the right panel is the bicoherence magnitude plot of the cloned recording.

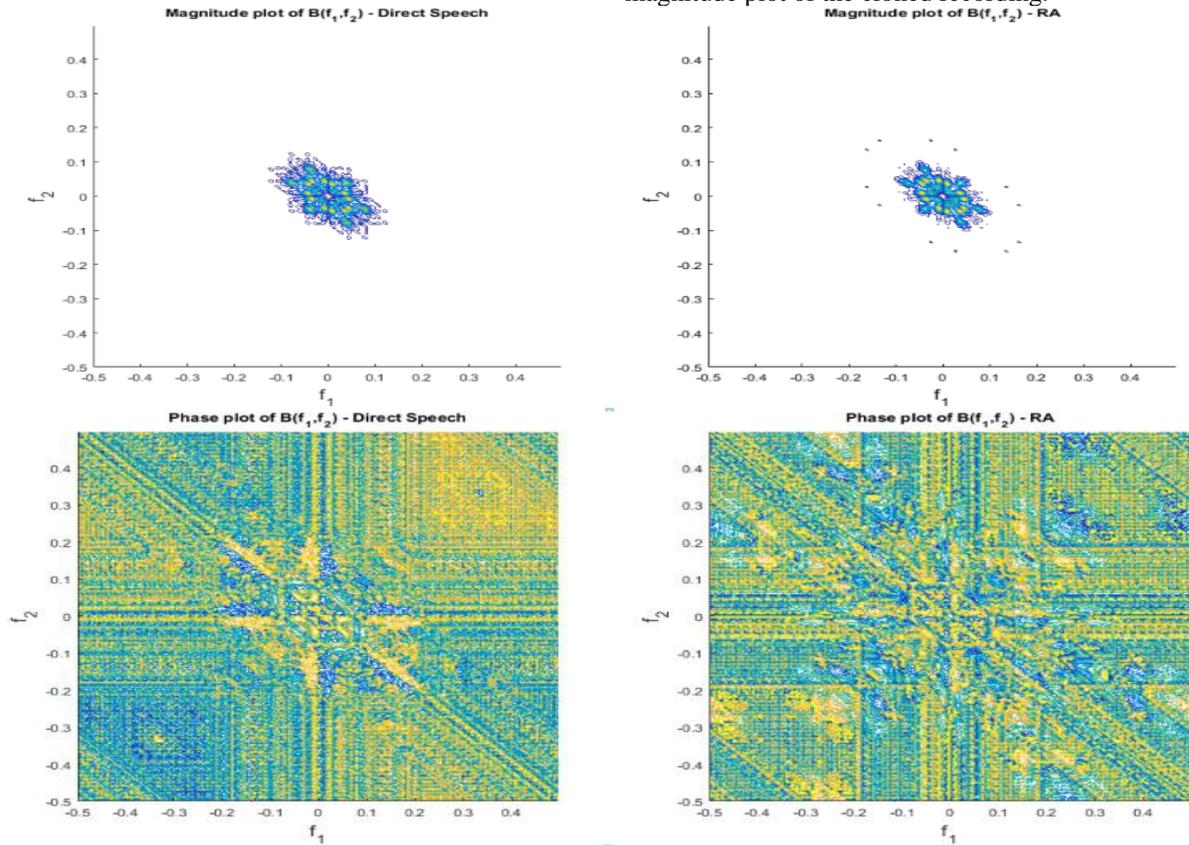

**Figure 3**: Shown in the top-left panel is the bicoherence magnitude plot for direct recording and top-right panel is the corresponding RA recoding. In the left-bottom panel is the phase bicoherence phase plot for the direct and RA recordings.

It can be observed from Fig. 3 that there is significant intermodulation distortion spread in the both bicoherence magnitude and phase spectra replay-audio recordings.

**3.2 Gaussianity test statistics and linearity test statistics-based detection**:

Gaussianity and linearity statistics tests can also be used to confirm non-Gaussianity and nonlinearity in a given stationary time series. It is reasonable to assume that bona-fide and relay attack speech signals are stationarity sequences. Moreover, bone-fide speech signal is also modeled as a non-Gaussian random sequence. The MSM processing chain of replay attack is expected to introduce nonlinearity in the resulting sequence. Let x(n) is non-Gaussian speech sequence and y(n) is linear non-Gaussian sequence of replay attack. How do we know that x[n] is non-Gaussian and y[n] is non-Gaussina and nonlinear? To achieve this goal, Hinich's non-skewness (also known as Gaussianity) and linearity tests [11] is used.

These tests relay on the fact that if the $3^{rd}$-order cumulants of a stationary process are zero, then its bicoherence is zero, and non-zero bicoherence implies that process is non-Gaussian. Moreover, if that the process is linear and non-Gaussian, then the bicoherence is a nonzero constant. Following binary hypothesis testing can be used for non-Gaussianity and nonlinearity detection:

H1 : the bispectrum of y(n) is nonzero and not constant;
H0 : the bispectrum of y(n) is nonzero and constant.

V.    EXPERIMENTAL RESULTS

For data collection, we used the Drop In Audio conferencing feature of the Amazon Alexa to replay voice recordings to the Google Home Speaker. We replayed a recording of "OK, Google, Turn on Office Lamp" via the audio-conferenced Amazon Alexa from another home. In second settings, we turned on and off the lights using Google Home remotely using Drop-in features of Alexa, as described in Figure 01. A total of 12 original recordings were replayed twice a) once at 1st point of replay to obtain set of twelve 1st order replay audios; b) The 1st order replay audios were replayed again at 2nd point to get another set of 2nd order replay cloned audios (See Fig.1). Next, we performed following three experiments.

**Experiment 1:** The goal of this experiment is to investigate impact of replay attack on bichorence magnitude and phase spectra. To this end, both the direct speech and RA recordings are segmented into frames of duration with a 50% overlapping factor. Bicoherence is estimated from each audio segment using the direct (fft-based) approach [8]. The bicoherence is estimated with the following parameter settings: 1) 1024-point segment length, 2) 1024-point FFT length, 3) 50% overlap, and 4) Rao-Gabr optimal window for frequency domain smoothing. Shown in Fig. 4 are the bispectrum magnitude and phase plots estimated from direct speech and corresponding RA recordings for third successful attack.

It can be observed from Fig. 4 that RA causes higher-order nonlinearity which is evident both in the magnitude and phase spectra. Similar, observations are made for other two other attacks.

**Experiment 2:** The goal of this experiment is to investigate impact of replay attack on Gaussianity test statistics and linearity test statistics. To this end, test statistics is calculated from both bona-fide and replay speech signals. To achieve this goal, Hinch's Gaussianity and linearity test statistics is calculated using *glstat function* available in the HOSA Matlab Toolbox [21], which can be used to estimate both Gaussianity test statistics and linearity test statistics. Frame-level Gaussianity test statistics and linearity test statistics are estimated from direct and RA attack recordings. It is observed that for all three RA recordings every non-silence frame failed Gaussianity and linearity test; whereas, all three direct recordings only less than 35% non-silence frames failed Gaussianity and linearity test. These findings confirm that RA introduce nonlinearity which can be used from RA detection.

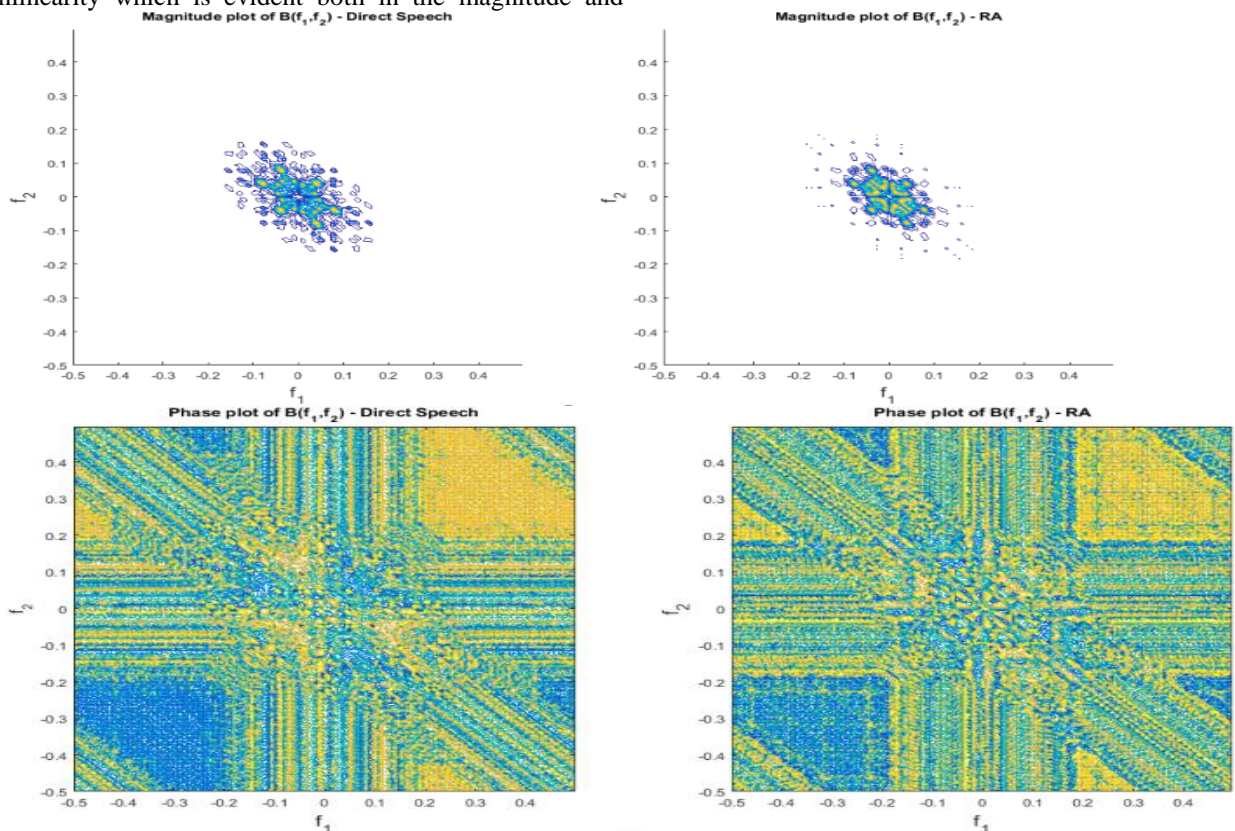

**Figure 4**: Shown in the top-left panel is the bicoherence magnitude plot for direct recording and top-right panel is the corresponding RA recoding for. In the left-bottom panel is the phase bicoherence phase plot for the direct and RA recordings.

**Experiment 3:** The goal of this experiment is the investigate impact of 2nd–order replay attack. It is expected that of 2nd–order would even introduce higher level of nonlinearity and stronger QPC. To validate this claim, a 2nd-order RA was recorded for all three attacks. Parametric QPC detection is applied on three direct speech, three 1st order RA, and three 2nd-order RA recordings. Shown in Fig 5 are the scatter graphs of frame-level QPC frequency locations estimated from all nine recordings.

It can be observed from Fig. 5 that RA causes shift of QPC peaks. This observation is consistent for all 1st – and 2nd – order replay attacks. Shift in QPC can be used for RA detection.

## Conclusion

This paper has demonstrated that the automatic speaker verification system used by Google home and Amazon Echo gadgets is vulnerable to replay attacks and thus all the skills and actions built on these platform, including many having critical financial data, could be exploited easily even by relatively less tech savvy impersonators. We performed vulnerability analysis and detection of replay attack using a) Drop-in features of Alexa by exploiting the replay attack at the Google Home, b) Alexa default services as well as skills developed using Alexa skills kit, c) Google home voice authentication.

Evaluation of proposed framework shows that HOSA-based features could be used to thwart replay attacks on Google Home and Amazon Alexa platform. More specifically, we demonstrated that RA causes higher-order nonlinearity which is evident both in the magnitude and phase spectra. Our results confirm our hypothesis that non-linearity introduced in RA can be used for its reliable detection

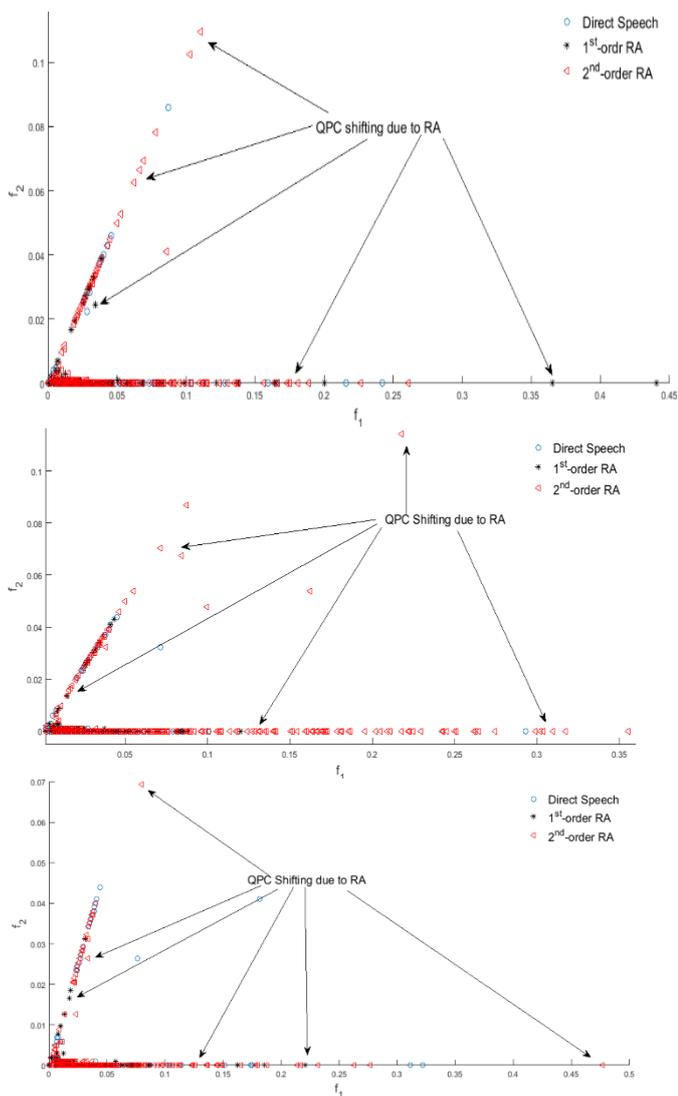

**Figure 5:** Scatter graph of frame-level QPC locations estimated from direct speech, $1^{st}$–order RA, and $2^{nd}$–order RA recordings.

ACKNOWLEDGMENT

This research is supported by National Science Foundation (NSF) under awards No. 1815724 and 1816019.